\documentclass[fleqn,usenatbib]{mnras}

\usepackage{amsmath}
\usepackage{txfonts}
\usepackage{graphicx}
\usepackage[T1]{fontenc}
\hypersetup{linkcolor=red} 

\makeatletter
\newcommand\footnoteref[1]{\protected@xdef\@thefnmark{\ref{#1}}\@footnotemark}
\makeatother

\def\equationautorefname~#1\null{equation~(#1)\null}

\title[Milky Way dynamical modelling with \textit{Gaia} data]{First \textit{Gaia} dynamical model of the Milky Way disc with six phase space coordinates: a test for galaxy dynamics}

\author[M. S. Nitschai et al.]{
Maria Selina Nitschai,$^{1,2,3}$ Michele Cappellari$^{3}$ and Nadine Neumayer$^{2}$\\
$^{1}$Heidelberg University, Seminarstra{\ss}e 2, 69117 Heidelberg, Germany\\
$^{2}$Max Planck Institute for Astronomy, K{\"o}nigstuhl 17, 69117 Heidelberg, Germany\\
$^{3}$Sub-department of Astrophysics, Department of Physics, University of Oxford, Denys Wilkinson Building, Keble Road, Oxford OX1 3RH, UK
}

\date{Accepted 2020 April 20. Received 2020 April 20; in original form 2019 September 11}
\pubyear{2020}

\begin{document}
\label{firstpage}
\pagerange{\pageref{firstpage}--\pageref{lastpage}}
\maketitle

\begin{abstract}
We construct the first comprehensive dynamical model for the high-quality subset of stellar kinematics of the Milky Way disc, with full 6D phase-space coordinates, provided by the \textit{Gaia} Data Release 2. We adopt an axisymmetric approximation and use an updated Jeans Anisotropic Modelling (JAM) method, which allows for a generic shape and radial orientation of the velocity ellipsoid, as indicated by the \textit{Gaia} data, to fit the mean velocities and all three components of the intrinsic velocity dispersion tensor. 
The Milky Way is the first galaxy for which all intrinsic phase space coordinates are available, and the kinematics are superior to the best integral-field kinematics of external galaxies. This situation removes the long-standing dynamical degeneracies and makes this the first dynamical model highly over-constrained by the kinematics. For these reasons, our ability to fit the data provides a fundamental test for both galaxy dynamics and the mass distribution in the Milky Way disc. We tightly constrain the volume average total density logarithmic slope, in the radial range 3.6--12~kpc, to be $\alpha_{\rm tot}=-2.149\pm 0.055$ and find that the dark halo slope must be significantly steeper than $\alpha_{\rm DM}=-1$ (NFW). The dark halo shape is close to spherical and its density is $\rho_{\rm DM}(R_\odot)=0.0115\pm0.0020$~M$_\odot$~pc$^{-3}$ ($0.437\pm0.076$~GeV~cm$^{-3}$), in agreement with previous estimates. The circular velocity at the solar position $v_{\rm circ}\left(R_{\sun}\right) = 236.5\pm 3.1$~km~s$^{-1}$  (including systematics) and its gently declining radial trends are also consistent with recent determinations.  
\end{abstract}

\begin{keywords}
Galaxy: disc -- Galaxy: kinematics and dynamics -- Galaxy: solar neighbourhood
\end{keywords}



\section{Introduction}

For decades astrophysicist have constructed dynamical models of external galaxies from observations of their unresolved stellar kinematics to study their masses as well as their orbital and density distributions \citep[see review by][]{Courteau2014}. The models were initially constrained by long-slit spectroscopy \citep[e.g.][]{vanDerMarel1991} and nowadays by integral-field spectroscopy \citep[see review by][]{cappellari16_rev}.

When using only projected line-of-sight kinematics and the first two velocity moments, there are well-known fundamental degeneracies between the mass density and the orbital distribution, or anisotropy, for spherical galaxies. This mass-anisotropy degeneracy \citep{Binney1982,Gerhard1993} led to the development of techniques to extract the elusive shape of the stellar line-of-sight velocity distribution from the galaxy spectra \citep[e.g.][]{Bender1990,vanderMare1993} together with dynamical modelling approaches that could use that information to break the degeneracy \citep[e.g.][]{Rix1997,vanderMarel1998,Gebhardt2000,Valluri2004,Cappellari2006}.

The mass-anisotropy degeneracy is less severe in axial symmetry as one can observe different views of the velocity ellipsoid along different polar angles on the sky. However, degeneracies must still be expected because the data cube is a three dimensional observable which cannot uniquely constrain the three-dimensional distribution function in addition to the galaxy density distribution \citep[see discussion in Sec.~3 of][]{Valluri2004}. Moreover, the surface-brightness deprojection is also strongly degenerate, unless the galaxy is edge-on \citep{Rybicki1987}. Dynamical degeneracies are indeed observed even with state-of-the-art dynamical models and data \citep{Krajnovic2005,deLorenzi2009}.

For external galaxies, heroic attempts were made to break the dynamical degeneracies using proper motion measurements in addition to line-of-sight kinematics \citep[e.g.][]{vanDeVen2006,Watkins2015}, but these useful proof-of-concept studies had to rely on data of relatively limited quality. Moreover, these studies are only possible for very few cases, where the full velocity information is available.

For the Milky Way there have been in the past years many surveys gathering kinematic information for stars all over the sky. For example the Geneva-Copenhagen Survey \citep{Nordstrom04}, RAVE \citep{RAVE06}, APOGEE \citep{APOGEE08,majewski17}, \textit{Gaia-ESO} \citep{GES2012} and many others, but the \textit{Gaia} mission \citep{Gaia_mission_16} is by far the largest one, measuring billions of sources. \textit{Gaia} provides us with the largest sample of three-dimensional kinematic information for stars that cover a large area on the sky.

With this information, all these long-standing dynamical degeneracy issues disappear for the kinematics of the Milky Way from the \textit{Gaia} mission, which are based on direct proper motion and radial velocities determinations for millions of individual stars: (i) one can obtain all six phase-space coordinates (three spatial coordinates and three velocities), making the dimension of the observable larger than that of the distribution function, thus allowing for extra parameters like the density to be uniquely constrained; (ii) one measures the true velocity moments, by direct summation, over many stars rather than having to infer them from integrated galaxy spectra; (iii) the stellar density is uniquely obtained without the need for a deprojection.

In this situation, unlike the case of {\rm every} galactic dynamical model for other galaxies that has been constructed in the past half a century, dimensional arguments alone already imply that there is no guarantee that even general models will be able to fit simultaneously all the components of the kinematics unless the model assumptions are sufficiently accurate. These data sets that provide full velocity information then allow us to make a fundamental physical test, namely to verify whether a model based on the Newtonian equations of motion\footnote{Relativistic corrections are unimportant here}, which was developed based on the motions on the scale of the Solar System, is able to accurately predict the average motions of the stars at the scale of our Galaxy, $10^8$ times larger.

In recent years there has been good progress in dynamical modelling of the Milky Way and a summary of different dynamical methods is given by \cite{rix13}. \cite{Bovy13} used the dynamical modeling based on action integrals suggested by \cite{Binney2010, Binney2011, Binney2012} where they use six-dimensional dynamical fitting with three-action-based distribution functions to fit abundance selected stellar populations from SEGUE. This machinery was improved and implemented in the \textsc{RoadMapping} code \citep{Trick16}, to recover the gravitational potential by fitting an orbit distribution function to stellar populations within the disc and it was applied to mock data investigating its capabilities. The local dark matter density was determined to be 0.0126 $q^{-0.89}$~M$_\odot$~pc$^{-3}$ by \cite{Piffl_2014}, were $q$ is the halo's axis ratio, using also an action based distribution function. They investigated the vertical mass density using kinematics from giant stars in RAVE and variations in $z$ from the number density determined by \cite{juric08}, fitting a distribution function to the kinematics and computing the vertical density profile until it fits the observed profile. Kinematic models based on Gaussian and Shu distribution functions were used by \cite{Sharma14} to constrain kinematic parameters of the Milky Way using RAVE data and full phase space data from the Geneva-Copenhagen Survey. \cite{Portail17} used the made-to-measure method to construct dynamical models of the bar region only, using kinematics from BRAVA, ARGOS and OGLE and recovering the bar pattern speed, and the stellar and dark matter mass distribution in this region. Another work by \cite{Robin17}, investigated a wide solar neighbourhood field using the Besan\c{c}on population synthesis model applied to RAVE and \textit{Gaia} DR1 data to reproduce the velocities, constrain the thin and thick disc dynamical evolution and determining the solar motion. In \cite{Hagen18} they combine TGAS and RAVE data to study the kinematics of red clump stars and derive the dark matter density in the solar neighbourhood. They apply axisymmetric Jeans equations and get $\rho_{\rm DM}(R_\odot, 0)$ = 0.018 $\pm$ 0.002~M$_\odot$~pc$^{-3}$. However, they also mention that the systematic errors are much larger and it is important to get accurate constrains on the stellar disc parameters.

Here we attempt to construct a first axisymmetric dynamical model of the Milky Way kinematics given by \textit{Gaia} DR2. We use the new spherically-aligned Jeans Anisotropic Method \citep[JAM$_{\rm sph}$,][]{Cappellari2019}, which allows for general axial ratios for all three components of the velocity ellipsoid and a spherical orientation, as indicated by the \textit{Gaia} data \citep{Hagen2019,Everall2019}. We want to test to what accuracy a relatively simple model can capture the richness of the \textit{Gaia} kinematics. We intentionally keep the model as simple as possible not to risk over-fitting kinematic features that may be due to non-equilibrium or non-axisymmetry (e.g. bar, spiral arms and warps) rather than tracers of the gravitational potential. We additionally provide a description of the mass density distribution (and circular velocity) of the Milky Way at radii $r\approx3.6$--12~kpc.

The outline of this paper is as follows: In Section~\ref{sec:data} we briefly present the \textit{Gaia} DR2 dataset and introduce the Jeans model, including its required components in Section~\ref{sec:methods}. We present the resulting JAM model and the Milky Way circular velocity curve in Section~\ref{sec:results}, before concluding in Section~\ref{sec:conclusion}.

\section{\textit{Gaia} Stellar Kinematic Data}\label{sec:data}

In April 2018 \textit{Gaia} had its second data release \citep[\textit{Gaia} DR2][]{Gaia2_18}, which contains the data collected during the first 22 months of its nominal mission lifetime.
This gives for the first time a high-precision parallax and proper motion catalogue for $\sim10^9$ sources, supplemented by precise and homogeneous multi-band all-sky photometry and a large (a few times $10^6$) radial velocity survey at the bright ($G < 13$) end \citep{Gaia2018}. 

Some early dynamical models used subsets of the DR2 \textit{Gaia} data to infer the shape and mass of the dark matter halo \citep{Posti2019,Watkins2019,Wegg2019}, the Galaxy's velocity curve \citep{eilers18} or its non-equilibrium features \citep{Antoja2018}. However, no study has yet attempted to construct a comprehensive model of the bulk of the \textit{Gaia} DR2, namely the few times $10^6$ of stars with high-quality full six-dimensional phase space coordinates. This is the goal of this paper.

We use kinematics derived with Bayesian distance estimates of \cite{Schoenrich19}. We assume as distance to the Galactic Centre $R_{\odot}=8.2$~kpc \citep{Abuter19}, a vertical displacement of the Sun from the midplane of $z_{\odot}=0.02$~kpc \citep{Joshi07} and as solar velocities in cylindrical Galactic coordinates $(U_{\odot}, V_{\odot}, W_{\odot})= (-11.1, 247.4, 7.2)$ km~s$^{-1}$ \citep[from][respectively]{Schoenrich10,Abuter19,Reid2009}. The giant stars are the main contribution in \textit{Gaia} at distances larger than 1~kpc from the Sun and can be measured out to large distance due to their brightness. For this reason they give the most homogeneous sub-sample over the area we want to probe. They are selected based on their absolute magnitude $M_G < 3.9$, intrinsic colour $(G_{BP}-G_{RP})_0 > 0.95$ and positive parallaxes with relative uncertainty $\varpi/\epsilon_{\varpi} > 5$, no other quality cuts were performed. This sample contains  about $1.98\times10^6$ stars and covers a volume with extreme cylindrical coordinates $3.65<R<12.02$~kpc, $-2.52<z<2.50$~kpc and $-15^\circ<\phi<15^\circ$, which is divided into $(R,z)$ cells of $200\times200$~pc. For each cell with at least 30 stars we calculated the median velocity and the velocity dispersion. The median uncertainties over our full sub-sample are $(\epsilon_{v_r}, \epsilon_{v_{\phi}}, \epsilon_{v_z})=(2.2, 1.3, 1.4)$~km~s$^{-1}$, while the median velocity error in each bin is always below 3~km~s$^{-1}$, making the uncertainties negligible for both velocity and velocity dispersion.

\section{Methods}\label{sec:methods}

\begin{figure*}
	\includegraphics[width=\textwidth]{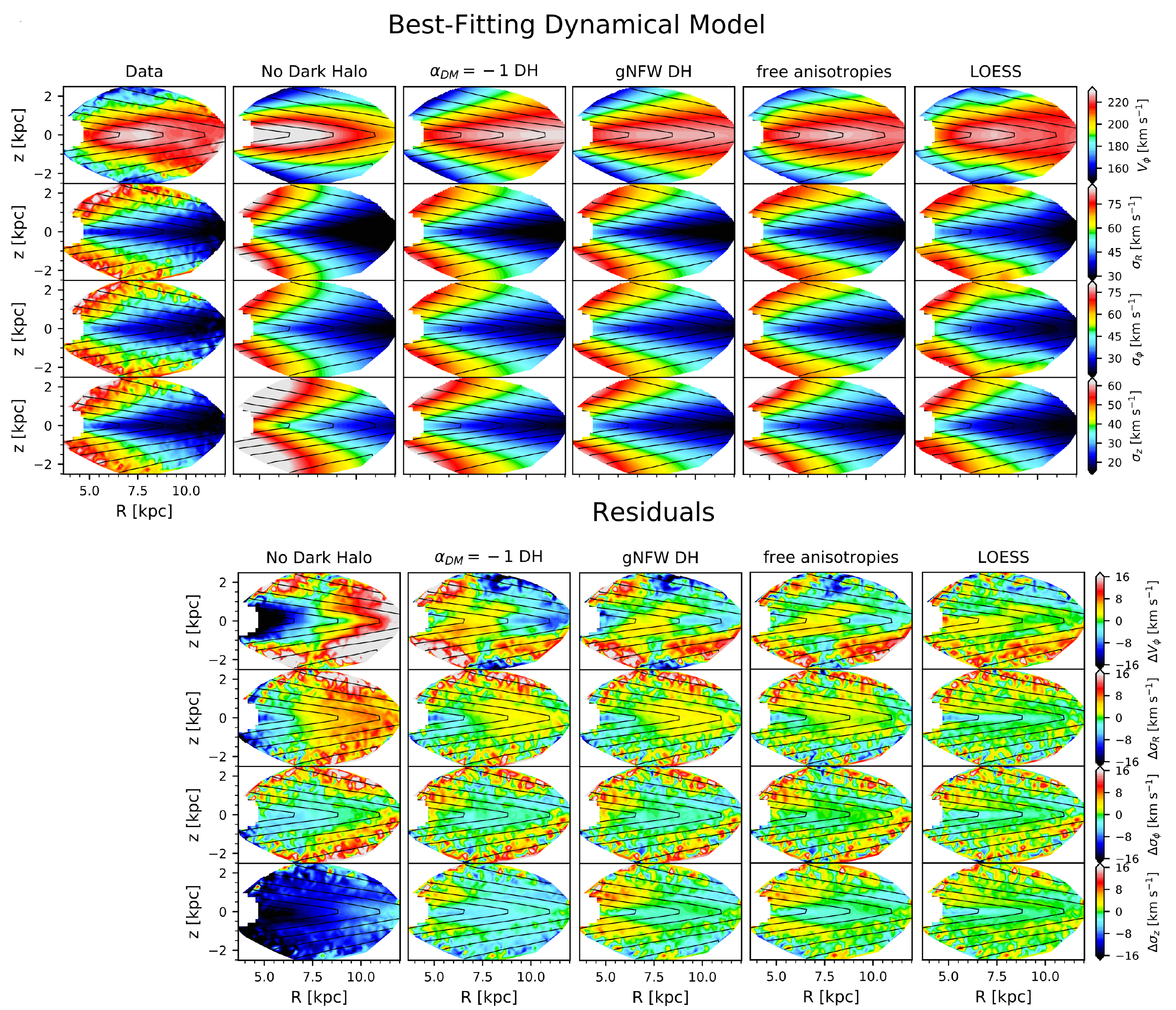}
     \caption{{\bf Data versus models.} From left to right: (i) \textit{Gaia} Data, (ii) the best-fitting JAM$_{\rm sph}$ model without dark matter, (iii) the best model with a `standard' Navarro, Frenk and White (NFW) dark matter profile, (iv) the best-fitting model with gNFW halo profile, with free inner logarithmic slope, and two free anisotropies (see text), (v) the best-fitting model with a different anisotropy for each MGE Gaussian and (vi) a symmetrized and LOESS-smoothed version of the \textit{Gaia} data. The latter is shown for reference and, given that it makes no other assumption than symmetry and small-scale smoothness, it essentially represents the best fit that one can expect with \emph{any} axisymmetric model. Below each model there are the corresponding residuals (${\rm data} - {\rm model}$). From top to bottom the rows show the mean azimuthal velocity ($v_{\phi}$), the velocity dispersion in radial ($\sigma_R$), azimuthal ($\sigma_{\phi}$) and vertical ($\sigma_z$) direction in Galactic cylindrical coordinates.}\label{im:JAM}
\end{figure*}

 \subsection{Model for the Milky Way stellar luminosity density}
 \label{sec:model_luminosity}\textbf{}
 
 The first component one needs to construct for a stellar dynamical model is the distribution of the tracer population, from which the kinematics was derived. Ideally, this would be extracted from the same \textit{Gaia} data set we use to derive the kinematics, but this kind of model is not (yet) available. 

Therefore, we use the Milky Way stellar distribution from \cite{juric08} derived from the number density of main sequence stars using data from the Sloan Digital Sky Survey (SDSS). However, we are ignoring the stellar halo component since it is quite uncertain and not as important since our data are mainly stars in the Milky Way disc rather than in the halo. The \cite{juric08} photometric model was derived with data that only cover the disc plane near the solar position and include only a smaller part of the Southern sky. Hence in our work it is extrapolated to describe the whole disc area and used for giant stars rather than main sequence stars from which it was derived. Nevertheless, it is still the best stellar distribution we have at the moment for the disc region and probably is still a good approximation for the giant stars too. In retrospect, the fact that our dynamical model, which has very little freedom, is able to fit the data well, support the idea that the adopted stellar model may be a reasonably good representation of our tracer distribution. In fact, any change in the tracer distribution directly translates into variations of the model kinematics.

The disc is decomposed into a sum of two exponential components, the thin and the thick disc. This allows for different scale lengths ($L_{\rm thin}, L_{\rm thick}$) and heights ($H_{\rm thin}, H_{\rm thick}$) for each component:
\begin{align} 
&\rho_{\rm D}(R, z, L, H) = \rho(R_\odot,0)\; \exp\left(- \frac{R-R_\odot}{L} - \frac{\left|z\right|}{H}\right)
\label{eq:disc},\\
&\rho_{\rm D}(R, z) = \rho_{D}(R, z, L_{\rm thin}, H_{\rm thin}) + f\times \rho_{\rm D}(R, z, L_{\rm thick}, H_{\rm thick}) 
\label{eq:disc components}
\end{align} 
where $\rho(R_\odot,0)$ is the number density of stars in the solar neighbourhood, and the parameter $f$ in \autoref{eq:disc components} is the thick disc normalization relative to the thin disc. Our adopted values are the `bias corrected' parameters \citep{juric08}: $f=0.12$; $(L_{\rm thin}, H_{\rm thin})=(2.6, 0.3)$~kpc and $(L_{\rm thick}, H_{\rm thick}) =(3.6, 0.9)$~kpc.

The parameter $\rho(R_\odot,0)$ was originally derived as a number density from star counts by \cite{juric08}, but since we want to obtain meaningful stellar mass-to-light values $(M_\ast/L)$ we normalize the density \citep{juric08} in such a way that the stellar luminosity density at the solar radius corresponds to the local luminosity density in \textit{V}-band $\rho_{L} = 0.056$~L$_{\odot V}$~pc$^{-3}$, which was derived by \cite{Flynn06} using the local luminosity function and the vertical structure of the disc. We have chosen the \textit{V}-band since the SDSS data, used to fit the disc model, and the \textit{Gaia} data are including the \textit{V}-band.

We further included the bulge, even though we have tested the sensitivity of our models to it and saw that it has a minimal effect on the disc area we are probing with the \textit{Gaia} data and therefore we will keep it fixed for our model. Even an extreme model without a bulge would not affect our conclusions. We use a bulge model of \cite{mcmillan17} which is an axisymmetric version of the one obtained from \textit{COBE/DIRBE} photometry \citep{Bissantz2002}:
\begin{align} 
\rho_{\rm b} &= \frac{\rho_{0,b}}{\left(1+m/r_0\right)^a}\; \exp\left(-\frac{m^2}{r_{\rm cut}^2}\right)
\label{eq:bulge},\\
m^2 &= R^2 + (z/q)^2.\label{eq:ell_rad}
\end{align}
We adopt the values of \cite{mcmillan17} $a = 1.8$, $r_0 = 0.075$~kpc, $r_{\rm cut} = 2.1$~kpc and an axis ratio of $q = 0.5$.
We normalize the bulge to our disc by ensuring it has the same bulge to disc ratio as the original model \citep{mcmillan17}, obtaining $\rho_{0,b} = 80.81$~L$_{\odot V}$~pc$^{-3}$.

Within 5~kpc from the Galactic Centre the bar dominates the kinematics \citep{Wegg15, Bovy19} but since our data are mostly not covering this area we are ignoring the bar in our stellar model. The small fraction of stars that might belong to the bar with radii smaller than 5~kpc should not have a significant effect on our model. Previous studies showed that one can measure reliable quantities outside the bar region of barred galaxies \citep{Lablanche12}, by symmetrizing the bar density as done here. 

For use with our model, we approximate the intrinsic density of the disc $+$ bulge stellar model with a multi-Gaussian expansion (MGE) \citep{Emsellem94, Mge2002}, by fitting a synthetic two-dimensional image using the MGE fitting method \citep{Mge2002} and \textsc{mgefit} software package\footnote{\label{note:mge}We use the Python version 5.0.2 of the \textsc{mgefit} package available from \url{https://pypi.org/project/mgefit/}}. MGE is a general and easy way to find the potential and solve the Jeans equations without having to modify the formalism for every different adopted parametrization. Most importantly, the method is sufficiently flexible to describe the photometry of multi-component galaxies in great detail. \cite{Cappellari2019} in Fig.~5 presents an example of the negligible ($\la1$ per cent) errors one can expect in the final Jeans kinematic predictions when approximating an analytic distribution with the MGE, as we do in this paper. This indicates that numerical approximation errors due to the limited terms in the expansion should be negligible with respect to systematic uncertainties in the data and model assumptions.

\subsection{Model for the dark matter and gas contributions}
\label{sec:model_dark}

The second component of a dynamical model is the mass density of the Galaxy, which includes not only the stellar component but also the gas distribution and the dark matter halo.

We describe the dark matter as a generalized Navarro, Frenk and White profile (gNFW, \citealt{Wyithe2001}), with variable inner slope $\alpha_{\rm DM}$ and axial ratio $q$:
\begin{equation}
\rho_{\rm DM} = \rho_{\rm s} \left(\frac{m}{r_{\rm s}}\right)^{\alpha_{\rm DM}}\left(\frac{1}{2}+\frac{1}{2}\frac{m}{r_{\rm s}}\right)^{-3-\alpha_{\rm DM}},
\label{eq:gNFW}
\end{equation}
with $m$ from \autoref{eq:ell_rad} and where the scale radius, $r_{\rm s}$, is also allowed to vary between 10 and 26~kpc, consistent with values found with the prediction of the halo mass-concentration relation $M_{200}-c_{200}$ \citep{NFW96,klypin11}, for a halo mass around $M_{200}\approx1.3\times10^{12}$~M$_\odot$ \citep{review16}, and with actual measurements  for the Milky Way \citep[e.g.][]{kafle14,mcmillan17,eilers18}.

If $\alpha_{\rm DM} = -1$, \autoref{eq:gNFW} is the classical Navaro, Frenk and White (NFW) dark matter profile \citep{NFW96}. This 1-dimensional profile is fitted with Gaussians using the \textsc{mge\_fit\_1d} procedure in the \textsc{mgefit} package (see footnote \ref{note:mge}), and appropriately made oblate/prolate, for use with the model.

In addition, we include the gas density, even though it is quite uncertain, again from the previous Milky Way model by \cite{mcmillan17}. We include the mass density of both the \ion{H}{I} and $H_2$ gas in the disc of the Milky Way (eq.~4 in \citealt{mcmillan17}). For simplicity of producing the MGE fit, we initially ignored the hole in the centre of this density profile, which is outside the range of our kinematics, by removing the term in the equation that describes it, so that the gas model only declines exponentially towards larger $r$. The gas density was modelled by creating an image which we fitted in 2-dim with the \textsc{mge\_fit\_sectors} routine in the \textsc{mgefit} package \citep{Mge2002} as we did for the stellar model. This gas MGE will be added to the potential density of the Milky Way but will be kept fixed to the quoted mass by \cite{mcmillan17} during the model fit, since we just want it as an estimate of the mass contribution from the gas in the disc region.

Formally, removing the hole changes the force/potential inside the volume we are probing and may affect our results, however we subsequently verified that the effect is negligible by also running a model including the central hole in the gas distribution. For an accurate MGE model for the gas with the hole, we had to proceed differently. We fitted eq.~(4) of \citealt{mcmillan17} in one dimension ($R$) along the equatorial plane ($z=0$) using the \textsc{mge\_fit\_1d} routine in the \textsc{mgefit} package \citep{Mge2002}, while allowing for negative Gaussians (\texttt{negative=True}). We then described the ${\rm sech}^2(0.5\times z/z_d)$ trend in $z$ with its best-fitting Gaussian $\exp(-0.195\times z^2/z_d^2)$, which provides an excellent approximation. The resulting 2-dim MGE model for the gas was then obtained by multiplying the 1-dim $R$-coordinate MGE with the $z$-Gaussian. Using this alternative 2-dim MGE gas model with a hole, we found that the total density slope only change well within the quoted errors. 

 \subsection{Bayesian JAM modelling} \label{sec:bayesian}

For the dynamical modelling we use a new solution of the axisymmetric Jeans equations under the assumption of a spherically-aligned velocity ellipsoid \citep{Cappellari2019}, which was shown to describing very accurately the dynamics of the \textit{Gaia} data both in the outer halo \citep{Wegg2019} and the disc region \citep{Hagen2019,Everall2019}. This JAM$_{\rm sph}$ method was developed specifically with the \textit{Gaia} data in mind. As an extreme test of the sensitivity of the model assumption on the orientation of the velocity ellipsoid, we also use for comparison the cylindrically-aligned JAM$_{\rm cyl}$ solution and the \textsc{jampy} package\footnote{\label{note:jam}We use the Python version 5.0.21 of the \textsc{jampy} package available from \url{https://pypi.org/project/jampy/}} \citep{JAM}. 

The JAM$_{\rm cyl}$ method was applied to model the integral-field stellar kinematics of large numbers of external galaxies (see review in \citealt{cappellari16_rev}) and has been extensively tested against N-body simulations \citep{Lablanche12, Li16} and in real galaxies against CO circular velocities \citep{leung18}, including many barred and non-perfectly-axisymmetric galaxies like the Milky Way. In both cases, it was found to recover unbiased density profiles, even more accurately than the more general Schwarzschild \citep{Schwarzschild79} approach (see e.g. Fig.~8 of \citealt{leung18}). A higher accuracy of JAM$_{\rm cyl}$ with respect to Schwarzschild was confirmed (see Fig.~6 of  by \citealt{Jin2019}) using the currently state-of-the-art Illustris cosmological N-body simulation  \citep{Vogelsberger2014}. Given its spherical alignment, the JAM$_{\rm sph}$ method should be even more accurate for the Milky Way.

JAM$_{\rm sph}$ is described in detail in \cite{Cappellari2019} and here we only summarize the key model assumptions. The Jeans equations with the velocity ellipsoid assumed to be aligned with the spherical coordinate system and the anisotropy defined as $\beta = 1- \sigma_{\theta}^2/\sigma_{r}^2$, become \citep[e.g.][eq. 1, 2]{Bacon83}:
  \begin{equation}
    \frac{\partial(\nu\overline{v_r^2})}{\partial r} +  \frac{(1+\beta)\nu\overline{v_r^2}-\nu\overline{v_{\phi}^2}}{r} = -\nu \frac{\partial\Phi}{\partial r},
    \label{eq:jeans1 sph}
  \end{equation}
  \begin{equation}
    (1-\beta)\frac{\partial(\nu\overline{v_r^2})}{\partial \theta} +  \frac{(1-\beta)\nu\overline{v_r^2}-\nu\overline{v_{\phi}^2}}{\tan \theta} = -\nu \frac{\partial\Phi}{\partial \theta}.
    \label{eq:jeans2 sph}
  \end{equation}
These can be combined to give a linear first order partial differential equation, for which standard textbook-like methods of solutions exist:
\begin{equation}
    \frac{(1-\beta)\tan \theta}{r}\frac{\partial(\nu\overline{v_{r}^2})}{\partial \theta} - \frac{2\beta\nu\overline{v_{r}^2}}{r} - \frac{\partial(\nu\overline{v_{r}^2})}{\partial r} = \Psi(r, \theta)
    \label{eq:comb jeans}
\end{equation}
where:
\begin{equation}
    \Psi(r, \theta) = v(r,\theta)\times \left(\frac{\partial \Phi}{\partial r} - \frac{\tan \theta}{r}\frac{\partial \Phi}{\partial \theta}\right).
\end{equation}
A detailed solution was given by \cite{Bacon83} and \cite{Bacon85}. \cite{Cappellari2019} specialized the Jeans solution to the case where both the density and the tracer distributions are described by the MGE parametrization and described an efficient and accurate numerical implementation, which we used in this work\footnote{It is included in the public \textsc{jampy} package from v6.0.}. The use of MGE allows one to spatially vary the anisotropy by assigning different anisotropies to the various Gaussian components of the MGE.

JAM$_{\rm sph}$ gives the solution of the Jeans equations for all three mean velocity components $(v_r, v_\theta, v_\phi)$ and the velocity dispersion in the three directions (eq. 52-54 of \citealt{Cappellari2019}). We project the spherical velocities into cylindrical coordinates $(v_R, v_\phi, v_z)$. However, given that the model assumes a steady state, $v_R$ and $v_z$ are identically zero and do not need to be explicitly fitted. This assumption is consistent with good accuracy with the \textit{Gaia} maps by \cite{Gaia2018} which also show these velocities to be small for the purpose of this study ($-15\la v_R\la15$~km~s$^{-1}$, $-10\la v_z\la10$~km~s$^{-1}$). 

The formal statistical errors in the \textit{Gaia} data are quite small and certainly much smaller e.g. than the actual deviations of the Milky Way kinematics (or most external galaxies) from the axisymmetry and steady-state assumptions. Moreover, the \textit{Gaia} data set contains a very large number of values. In these situations, the formal statistical uncertainties become meaningless as the uncertainties become entirely dominated by systematics. This is a common issue also for the dynamical modelling of high-$S/N$ integral-field kinematics. To approximately account for systematic errors in the data and approximations in the model assumptions we follow the approach of Sec.~3.2 of a previous modelling paper by \cite{vandenBosch2009}, as modified for Bayesian analysis in Sec.~6.1 of \cite{mitzkus17}. We stress the fact that the method is {\em not} statistically rigorous, as one should expect from the fact that it tries to deal with systematic and not statistical uncertainties. However it was found to work well in practice, in a number of cases. The key idea of this approximate approach to deal with systematic uncertainties in the data is to {\em assume} that they produce a comparable contribution to the final uncertainty on the model parameters as the statistical uncertainties in the data. This implies that, instead of adopting the standard statistical confidence levels based on $\Delta\chi^2$ (e.g. $\Delta\chi^2=1$ for the 1$\sigma$ uncertainty for one DOF) one should allow variations in the $\chi^2$ on the order of its standard deviation $\Delta\chi^2=\sqrt{2(N-M)}\approx\sqrt{2N}$, with $N=817$ the number of data points fitted and $M$ the number of model parameters. 

In practice, we proceed as follows:  First, after an initial fit, we adopt for the kinematics a constant error of $\epsilon=5.7$~km~s$^{-1}$, for all kinematic components, to give $\chi^2/{\rm DOF} = 1$ for the best-fitting model. 
To include this uncertainty in the Markov chain Monte Carlo (MCMC) sampling, the error needs to be increased in such a way that a `misfit' with a $\Delta \chi^2=\sqrt{2N}$ using the original errors results in a fit with $\Delta \chi^2=1$ with the updated, scaled errors. Hence, we multiply the original error $\epsilon$ = 5.7~km~s$^{-1}$ by $\left(2N\right)^{0.25}$, which is equivalent to changing the $1\sigma$ confidence level from $\Delta\chi^2=1$ to $\Delta\chi^2=\sqrt{2N}$.

JAM$_{\rm sph}$ or JAM$_{\rm cyl}$ uses as fixed input the density of the tracer population \citep{juric08} and a model for the gas density \citep{mcmillan17}. Our standard model has 9 free parameters: (i) the velocity dispersion ratios or anisotropies ($\sigma_{\theta}/\sigma_r$\footnote{For JAM$_{\rm cyl}$ this is $\sigma_z/\sigma_R$ \label{note:beta}} and $\sigma_{\phi}/\sigma_r$\footnote{For JAM$_{\rm cyl}$ this is $\sigma_{\phi}/\sigma_R$ \label{note:gamma}}) for both the flattest ($q_{\rm MGE}<0.2$) Gaussian components (subscript 1) of the MGE and the rest (subscript 2), to allow for some of the observed clear spatial variation of the anisotropy, while keeping the model as simple as possible. Our results are only weakly sensitive to different choices for separating the Gaussians with different anisotropies (see \autoref{sec: atot dens}); (ii) the inner logarithmic slope of the gNFW profile ($\alpha_{\rm DM}$); (iii) the dark matter fraction $f_{\rm DM}$ within a sphere of radius $R_\odot$; (iv) the mass-to-light ratio $M_\ast/L_V$ of the stellar component in the \textit{V}-band; (v) the axial ratio $q$ and (vi) the scale radius ($r_{\rm s}$) of the dark matter profile. 

Additionally, we also consider for comparison a model where we allow each of the 18 Gaussians in the MGE model for the stellar luminosity-density distribution to have a different anisotropy $\sigma_{\theta}/\sigma_r$ and $\sigma_{\phi}/\sigma_r$. This model has 41 free parameters.

The Bayesian modelling was performed using the \textsc{AdaMet}\footnote{We use the Python version 2.0.7 of the \textsc{AdaMet} package available from  \url{https://pypi.org/project/adamet/}} package of \cite{cappellari13}, which implements the Adaptive Metropolis algorithm by \cite{haario01}. This is used to estimate the posterior distribution, as in standard MCMC methods \citep{gelman2013bayesian}, to get the confidence levels of the best-fitting parameters and to show the relations between the different parameters. We adopted constant priors on all parameters, in such a way that the probability of the model, given the data, is just proportional to the likelihood $P({\rm data}|{\rm model})\propto\exp(-\chi^2/2)$.

\begin{table*}
	\centering
	\caption{Median parameters and 68 per cent (1$\sigma$) confidence intervals for the spherically or cylindrically-aligned JAM models and the $\chi^2$ values for the best-fitting models} \label{ta:JAM}
	\scalebox{.88}{
	\begin{tabular}{l|cccccc|c}
	    \hline
		Parameter 
		\vline
		& \multicolumn{6}{c}{Spherically-aligned JAM$_{\rm sph}$}
		\vline
		& Cylindrically-aligned JAM$_{\rm cyl}$\\
		 & No Dark Halo & NFW Dark Halo & gNFW Dark Halo & $-15^\circ<\phi<0^\circ$ & $0^\circ<\phi<15^\circ$ & free anisotropies & gNFW Dark Halo \\
		\hline
		(1) $\alpha_{\rm DM}$ & - & -1 & -1.53 $\pm$ 0.12 & -1.52 $\pm$ 0.13 & -1.50 $\pm$ 0.12  & -1.54 $\pm$ 0.14 & -1.48 $\pm$ 0.13\\
		(2) $f_{\rm DM}$ & - & 0.73 $\pm$ 0.05 & 0.86 $\pm$ 0.06 & 0.85 $\pm$ 0.06 & 0.84 $\pm$ 0.05  & 0.90 $\pm$ 0.06 & 0.91 $\pm$ 0.05\\
		(3) $ r_{\rm s} $ [kpc] & - & 10.8 $\pm$ 1.0& 16.8 $\pm$ 5.4 & 17.2 $\pm$ 5.3 & 17.1 $\pm$ 5.4 & 17.6 $\pm$ 4.9 &  16.2 $\pm$ 5.2 \\
		(4) $q_{\rm DM}$ & - & 1.16 $\pm$ 0.25 & 1.14 $\pm$ 0.21 & 1.12 $\pm$ 0.23 & 1.33 $\pm$ 0.31  & 1.03 $\pm$ 0.22 & 1.40 $\pm$ 0.35\\
        (5) $(\sigma_{\theta}/\sigma_r)_{1}$ & 0.97 $\pm$ 0.14 & 0.67 $\pm$ 0.05 & 0.62 $\pm$ 0.05 & 0.61 $\pm$ 0.05 & 0.60 $\pm$ 0.05 & - & 0.63 $\pm$ 0.04~\ref{note:beta}\\
	    (6) $(\sigma_{\theta}/\sigma_r)_{2}$ & 0.82 $\pm$ 0.11 & 0.75 $\pm$ 0.08 & 0.71 $\pm$ 0.09 & 0.76 $\pm$ 0.10 & 0.73 $\pm$ 0.11 & - & 0.83 $\pm$ 0.06~\ref{note:beta}\\
		(7) $(\sigma_\phi/\sigma_r)_{1}$ & 0.87 $\pm$ 0.20 & 0.79 $\pm$ 0.08 & 0.80 $\pm$ 0.08 & 0.78 $\pm$ 0.09 & 0.77 $\pm$ 0.08 & - & 0.80 $\pm$ 0.07~\ref{note:gamma}\\
		(8) $(\sigma_\phi/\sigma_r)_{2}$ & 0.87 $\pm$ 0.20& 0.94 $\pm$ 0.12 & 0.93 $\pm$ 0.12 & 0.95 $\pm$ 0.15 & 0.97 $\pm$ 0.14  & - & 0.96 $\pm$ 0.13 ~\ref{note:gamma}\\
		(9) $(M_\ast/L)_V$ & 1.82 $\pm$ 0.03 & 0.58 $\pm$ 0.12 & 0.30 $\pm$ 0.13 & 0.34 $\pm$ 0.13& 0.38 $\pm$ 0.13  & 0.21 $\pm$ 0.14 & 0.26 $\pm$ 0.12\\
		(10) $\chi^2$ & 10611 & 3672& 3251 & 3078 & 2816 & 2698& 3460\\\
		(11) $\chi^2_{\rm JAM}/\chi^2_{\rm LOESS}$ & 4.82 & 1.67 & 1.48 & 1.41 & 1.44 & 1.23 & 1.57\\
		(12) $\left(\chi_{\rm JAM}/\chi_{\rm LOESS}\right)^2_{V_\phi}$ & 6.39 & 2.37 & 1.98 & 1.91 & 1.89 & 1.38 & 2.06\\
		(13) $\left(\chi_{\rm JAM}/\chi_{\rm LOESS}\right)^2_{\sigma_R}$ & 2.52 & 1.29 & 1.22 & 1.20 & 1.32 & 1.14 & 1.30\\
		(14) $\left(\chi_{\rm JAM}/\chi_{\rm LOESS}\right)^2_{\sigma_\phi}$ &1.61 & 1.21 & 1.29 & 1.11 & 1.22 & 1.12 & 1.20\\
		(15) $\left(\chi_{\rm JAM}/\chi_{\rm LOESS}\right)^2_{\sigma_z}$ & 14.03 & 1.87 & 1.47 & 1.32 & 1.49 & 1.27 & 1.87\\
		\hline
	\end{tabular}}
	 \vspace{1ex}
	 
     \raggedright {\footnotesize {\it Notes:} Row (1): the inner logarithmic slope of the gNFW profile; Row (2): the dark matter fraction within a sphere of radius $R_\odot=8.2$ kpc; Row (3): the scale radius and Row (4): the axial ratio of the gNFW. Row (5) and (6): the velocity dispersion ratio $(\sigma_{\theta}/\sigma_r)_{1}$ and $(\sigma_{\theta}/\sigma_r)_{2}$. Similar row (7) and (8): the velocity dispersion ratio $(\sigma_{\phi}/\sigma_r)_{1}$ and $(\sigma_{\phi}/\sigma_r)_{2}$. Row (9): the mass to light ratio. Row (10): the $\chi^2$ of the best-fitting models and row (11): quality of the best-fitting models, with $\chi^2_{\rm JAM}$ measured from the JAM models and  $\chi^2_{\rm LOESS}$ from the symmetrized data. Row (12), (13), (14) and (15): the quality of the best-fitting models for each individual velocity component ($v_{\phi}, \sigma_R, \sigma_{\phi}, \sigma_z$).}
\end{table*}

\section{Results}\label{sec:results}

\subsection{JAM fit to the \textit{Gaia} data}

Formally, our standard model has 9 free parameters, however, most of them are either directly constrained by the data or irrelevant and marginalized over. We include these 9 parameters just not to risk underestimating the model uncertainties. The four ratios between the different components of the velocity dispersion can be directly measured from the maps, while the $M_\ast/L$ has an almost one-to-one correspondence to the dark matter fraction $f_{\rm DM}$. The halo scale length $r_s$ is totally unconstrained by the \textit{Gaia} data and is degenerate with $\alpha_{\rm DM}$: their combination simply defines the density total slope. While the halo axial ratio $q$ is consistent with a spherical shape. This effectively leaves to the model the freedom to vary only the two parameters $f_{\rm DM}$ and $\alpha_{\rm DM}$, which describe the dark matter halo, to fit a set of four two-dimensional kinematic maps!

The best-fitting standard gNFW model for the velocity maps is shown in \autoref{im:JAM}. We also show for comparison the model without dark matter and the best-fitting model with a `standard' Navarro, Frenk and White (NFW) $\alpha_{\rm DM}=-1$ profile \citep{NFW96}. Moreover, we also show the model with free anisotropy for each of the 18 Gaussians of the MGE. The median parameters from the posterior distribution for all four models are listed in \autoref{ta:JAM} together with the 1$\sigma$ uncertainties, defined as half of the intervals enclosing 68 per cent of the posterior values, marginalized over the other parameters.

In addition, in \autoref{im:JAM} we show for reference a version of the {\em Gaia} kinematic data that was symmetrized with respect to the equatorial plane and LOESS smoothed following \citet{Cappellari15}. In practice, first we use \textsc{symmetrize\_velfield}\footnote{\label{note:sym} This routine is included in the \textsc{plotbin} package available from \url{https://pypi.org/project/plotbin/}} on our {\em Gaia} data to generate a symmetric (`axisymmetric') version with respect to $z=0$ and then we use the \textsc{loess\_2d}\footnote{\label{note:loess} We use the Python version 2.0.11 of the \textsc{loess} package available from \url{https://pypi.org/project/loess/}} routine (with \texttt{frac=0.05}) of \cite{loess_2013}, which implements the two-dimensional LOESS algorithm of \cite{Cleveland88}. This gives a LOESS smoothed estimate of the kinematics at the sets of coordinates ($R$, $z$) for the symmetrized data.

We use the symmetrized and smoothed {\em Gaia} data as benchmark to quantify the quality of our JAM fits in \autoref{ta:JAM}. For this, we list the quantity $\chi^2_{\rm JAM}/\chi^2_{\rm LOESS}$, following \citet{Cappellari15}, for each of the best-fitting models. This quantity has the advantage over the usual $\chi^2$ that it approximates the $\chi^2/{\rm DOF}$ but is insensitive to the normalization of the kinematic uncertainties.

To partially estimate the effect on the model parameters of the non equilibrium in the kinematics of the Milky Way we also fitted two models to two independent subsets of the \textit{Gaia} data extracted from two azimuthal sectors ($-15^\circ<\phi<0^\circ$ and $0^\circ<\phi<15^\circ$) of our data. The results for the two sectors (\autoref{ta:JAM}) are consistent with each other and with the main model. As an extreme test of the sensitivity of our model to the assumptions on the orientation of the velocity ellipsoid, we also have fit a model (JAM$_{\rm cyl}$) which assumes a cylindrically-aligned velocity ellipsoid \citep{JAM}. The JAM$_{\rm sph}$ model gives a slightly better fit to the data than JAM$_{\rm cyl}$, consistently with the finding that the Milky Way velocity ellipsoid is nearly spherical aligned \citep{Hagen2019,Everall2019}. However, the difference between the parameters inferred by two, JAM$_{\rm sph}$ and JAM$_{\rm cyl}$, models is minimal (\autoref{ta:JAM}) as the two solutions are not very different in the disc plane.

Three results can be inferred from \autoref{im:JAM}: (i) The most striking and important is how well this simple equilibrium model is able to capture the average \textit{Gaia} kinematics. Here the main model residuals appear due to the known $\la15$ per cent non-equilibrium and non-axisymmetry features \citep{Gaia2018}, which cannot be described by an equilibrium model. In fact, most of the model residuals seem associated to the spiral-wave pattern visible in the face-on view of the kinematics of the Milky Way disc in Fig. 10 of \cite{Gaia2018}. The fact that most of the residuals are due to non-axisymmetry is also clearly visible by comparing the residuals of the best-fitting model to those of the symmetrized data (\autoref{im:JAM}): the largest JAM residuals are in the $V_\phi$ components, but most of those also stand out with respect to the symmetrized {\em Gaia} data. The same is true for the major deviations in the JAM fits to the other components. This good agreement is further quantified by the $\chi^2_{\rm JAM}/\chi^2_{\rm LOESS}$ ratio in \autoref{ta:JAM}, which shows values quite close to one, especially when allowing for `free anisotropies'. (ii) It is clear that a model without dark matter completely fails to describe the observations, (iii) moreover, a standard NFW dark matter profile does not provide an equally good fit as the best-fitting one, with our median dark halo slope $\alpha_{\rm DM}=-1.53\pm0.12$. In particular, an NFW halo systematically over-estimates $\sigma_z$ and the radial gradient in $v_\phi$. The halo slope lies in the range expected from simple predictions for halo contractions \citep{Gnedin2004} for samples of real galaxies (e.g. Fig.~2 in \citealt{cappellari13}). This is consistent with previous work indicating a steeper slope than the standard NFW is needed in the disc region of the Milky Way around the solar region \citep{Portail17, Cole17}. 

The anisotropies of the data and model are in approximate agreement, although the model is by construction smoother than the data and does not try to reproduce the sharp anisotropy variations and asymmetries. These are unlikely to contain information on the gravitational potential, but are instead due to non-equilibrium and non-axisymmetry. With a more general model like \citet{Schwarzschild79} method one could fit every detail of the data, down to the noise. However, a better fit will not necessarily constitute an improvement in the recovered density, because there is a risk of interpreting non-equilibrium or non-axisymmetry features as tracers of an axisymmetric equilibrium model. Tests on real galaxies \citep{leung18} and simulations \citep{Jin2019} indicate that Schwarzschild models provides a less accurate recovery of the true density than JAM in this situation. Thus, we intentionally kept the standard model simple, not to risk over-interpreting features in the data that are not real or not tracer of an equilibrium potential.

Our model has a total density at the solar position of $\rho_{\rm tot}(R_\odot)=0.0640\pm0.0043$~M$_\odot$~pc$^{-3}$ and a dark matter density of $\rho_{\rm DM}(R_\odot)=0.0115\pm0.0020$~M$_\odot$~pc$^{-3}$, corresponding to a dark matter energy density of $0.437\pm0.076$~GeV~cm$^{-3}$ where the uncertainties include systematics and are not purely statistical. These values are broadly consistent with previous determinations \citep{McKee2015,mcmillan17}.

\begin{figure*}
 \includegraphics[width=1.6\columnwidth]{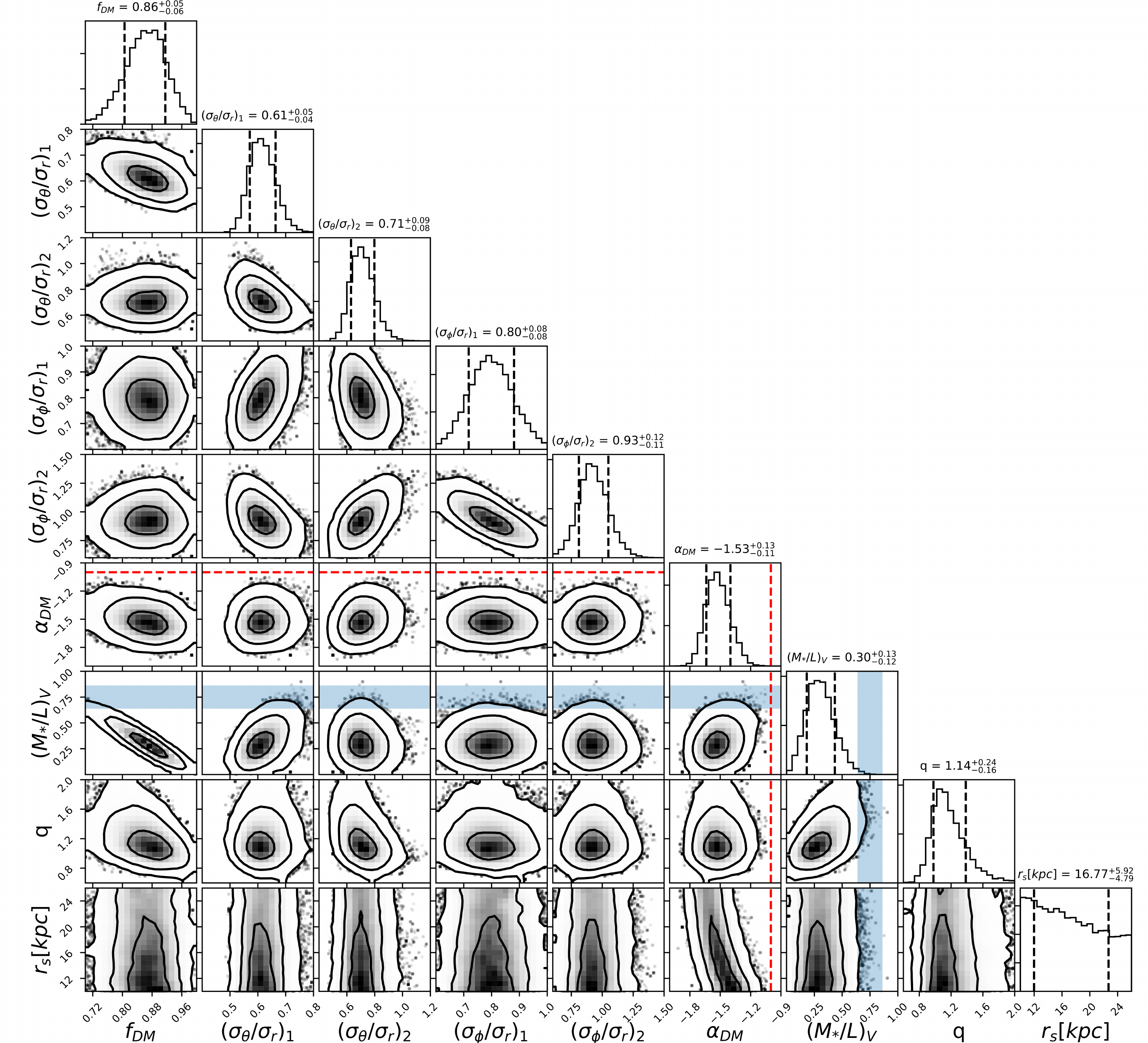}
 \caption{\textbf{gNFW parameters corner plot.} Each panel shows the posterior probability distribution  marginalized over two dimensions (contours) and one dimension (histograms). The parameters are (i) the dark matter fraction $f_{\rm DM}$ inside a sphere of radius $R_\odot=8.2$~kpc; (ii) the velocity dispersion ratios $\left[(\sigma_{\theta}/\sigma_r)_1,(\sigma_{\phi}/\sigma_r)_1\right]$ of the Gaussians flatter than $q_{\rm MGE}=b/a=0.2$ and the rest $\left[(\sigma_{\theta}/\sigma_r)_2,(\sigma_{\phi}/\sigma_r)_2\right]$; (iii) the inner dark matter halo logarithmic slope $\alpha_{\rm DM}$, (iv) the stellar mass-to-light ratio $(M_\ast/L)_{V}$; (v) the axial ratio $q$ of the dark matter halo and (vi) the scale radius $r_{s}$ for the dark matter halo. The thick contours represent the 1, 2 and 3$\sigma$ confidence levels for one degree of freedom. The red dashed line marks the `standard' $\alpha_{\rm DM}=-1$ slope of the Navarro, Frenk and White (NFW) \citep{NFW96} dark matter profile and the blue shaded band indicates a mass-to-light ratio estimate from stellar counts \citep{Flynn06}. The numbers with errors on top of each plot are the median (black dashed lines) and 16th and 84th percentiles of the posterior for each parameter, marginalized over the other parameters.}
\label{im:corner}
\end{figure*}

The posterior distribution of the parameters for the gNFW model are shown in \autoref{im:corner}. This figure also shows the expected covariance between the mass-to-light ratio, the dark matter fraction and the inner slope of the halo: models with steeper (more negative) $\alpha_{\rm DM}$ have lower $(M_\ast/L)_V$ ratio and higher dark-matter fraction $f_{\rm DM}$. This distribution also shows that an NFW profile (red dashed line in \autoref{im:corner}) is outside the 3$\sigma$ confidence level. In addition, as comparison the blue shaded region indicates the stellar mass-to-light ratio from stellar counts 0.75 $M_{\odot}/L_{\odot V}$ (15 per cent uncertainty at 1$\sigma$ level from \citealt{Flynn06}). This shows that our model $(M_\ast/L)_V$ is consistent with the stellar-counts determination within the errors. Our model strongly excludes flattened haloes and favours a nearly round one, consistent with recent \textit{Gaia} results \citep{Wegg2019}.

\subsection{Total density profile}\label{sec: atot dens}

\begin{figure}
	\centering
	\includegraphics[width=\columnwidth]{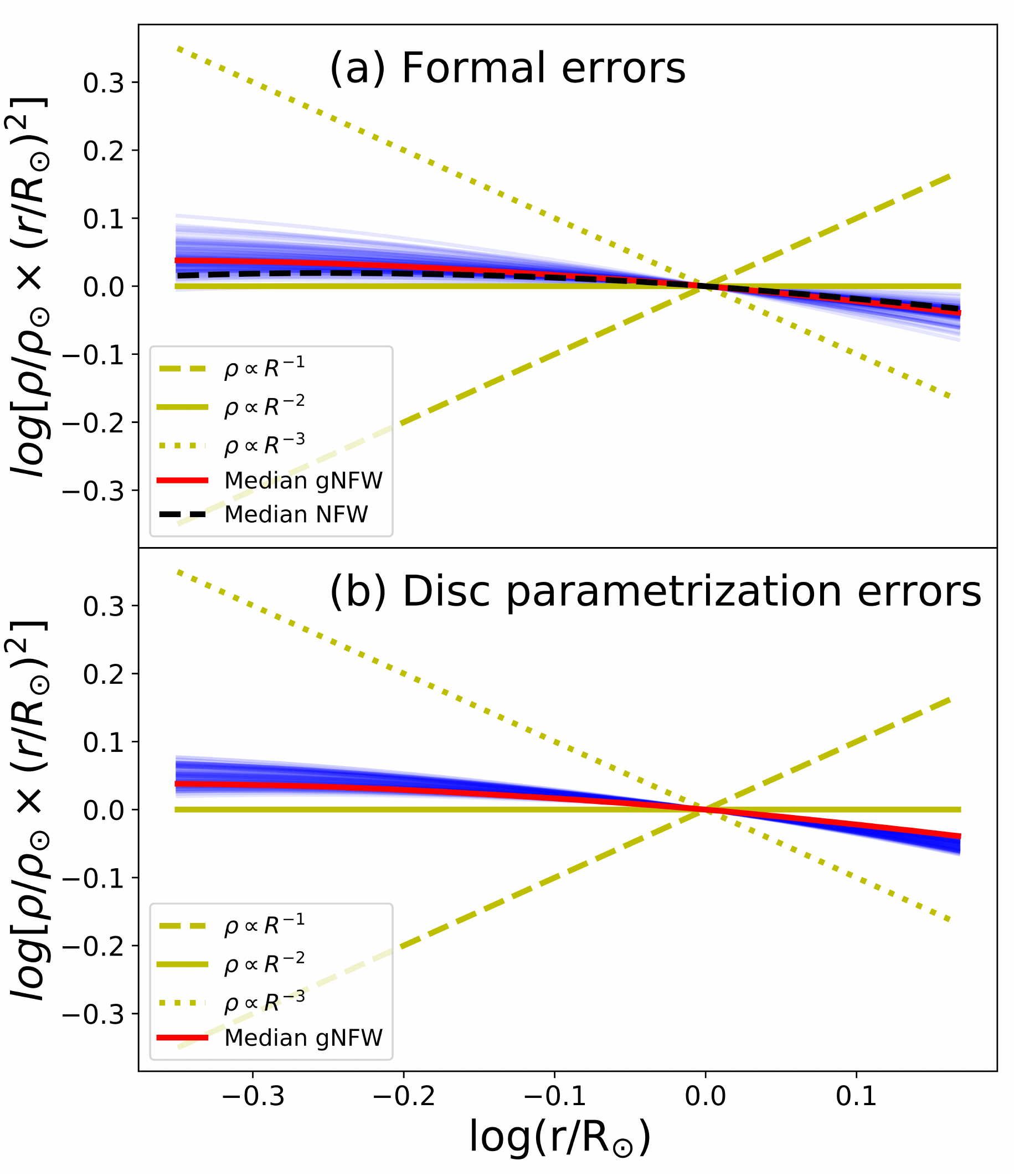}
	\caption{{\bf Milky Way total density profile.} (a) The blue shaded region shows 100  realizations of the density profile from the model posterior of \autoref{im:corner}. The black line is the median density for the NFW $\alpha_{\rm DM}=-1$ model. (b) Density profiles obtained by randomly varying the assumed parametrization for the stellar disc \citep{juric08} within the quoted errors and re-running the whole model fitting procedure.}
	\label{im:density}
\end{figure}

Having found our best-fitting model we can find the total density profile, using the \textsc{mge\_radial\_density} procedure in the \textsc{jampy} package (see footnote \ref{note:jam}).

The {\em total}-density distribution is the quantity that the dynamical models directly measure. This is very tightly constrained by the \textit{Gaia} data and it is shown in \autoref{im:density}. From the posterior of the models we measure a total density mean logarithmic slope of $\alpha_{\rm tot}\equiv\Delta\log\rho_{\rm tot}/\Delta\log r = -2.149\pm 0.055$ in the radial range 3.6--12~kpc of the data. These radii correspond to about 0.9--2.9 half-light radii $R_{\rm e}^{\rm max}$, for our measured\footnote{Using \textsc{mge\_half\_light\_isophote} in the JAM package.} $R_{\rm e}^{\rm max}\approx4.122$~kpc. The measured total-density slope is consistent with the `universal' slope $\alpha_{\rm tot}\approx-2.19\pm0.03$ inferred at comparable radii on early-type disc galaxies \citep{Cappellari15} with effective velocity dispersion not smaller than the Milky Way's value \citep{Kormendy2013} $\sigma_{\rm e}=105\pm20$~km~s$^{-1}$. 

We also investigate how much the uncertainties of the parametrization of the Milky Way stellar tracer distribution affect our JAM$_{\rm sph}$ results. For this, we vary the disc parameters of \cite{juric08} within the quoted errors and then redo the JAM$_{\rm sph}$ fit using the \textsc{capfit} least-squares program\footnote{This is part of the Python package \textsc{pPXF} by \cite{cappellari17} available here \url{https://pypi.org/project/ppxf/}}. This is repeated for 100 different disc models, for random values of the scale heights, scale lengths and the fraction of the thick disc normalization, within the given 20 per cent 1$\sigma$ uncertainties \citep{juric08}. The result for the total density can be seen in panel (b) of \autoref{im:density}. Here the blue lines indicate the systematic error due to the Milky Way stellar tracer model. This gives us an alternative estimate of the value and a systematic error $\alpha_{\rm tot}\equiv\Delta\log\rho_{\rm tot}/\Delta\log r= -2.194 \pm 0.044$ and the logarithmic slope of the gNFW profile is $\alpha_{\rm DM} = -1.455 \pm 0.055$, which is comparable to the previous estimate.

During this testing of these systematic errors, we saw that higher values for the scale height and length of our disc models seem to give JAM models that agree even better with our data. We leave further exploration of this aspect to a subsequent study.

Our model results depend on some arbitrary choices, like the number of Gaussians used in the MGE fit, and the axial ratio $q_{\rm MGE}$ at which we separate Gaussians with different anisotropies. We tested the sensitivity of the most robust quantity from the model, the total density slope, to these choices. If we force the MGE fit to have 35 Gaussians, instead of the 18 of our standard model, and perform a least-squares fit, we obtain a slope for the total density of $\alpha_{\rm tot} = -2.203$, which agrees within the quoted errors with the least-squares fit for our standard model which has $\alpha_{\rm tot} = -2.185$ (this is slightly different from the median value from the model posterior quoted previously in this Section). Additionally, if we allow four possible anisotropies, separated both in radius (smaller or larger than one effective radii) and flattening  at $q_{\rm MGE} = 0.3$ (instead of our standard $q_{\rm MGE} = 0.2$) we get $\alpha_{\rm tot} = -2.075$. So even with this different choice to separate the anisotropies, the result is still consistent at the level of the quoted errors.

\subsection{Circular velocity curve}

\begin{figure}
    \centering
    \includegraphics[width=\columnwidth]{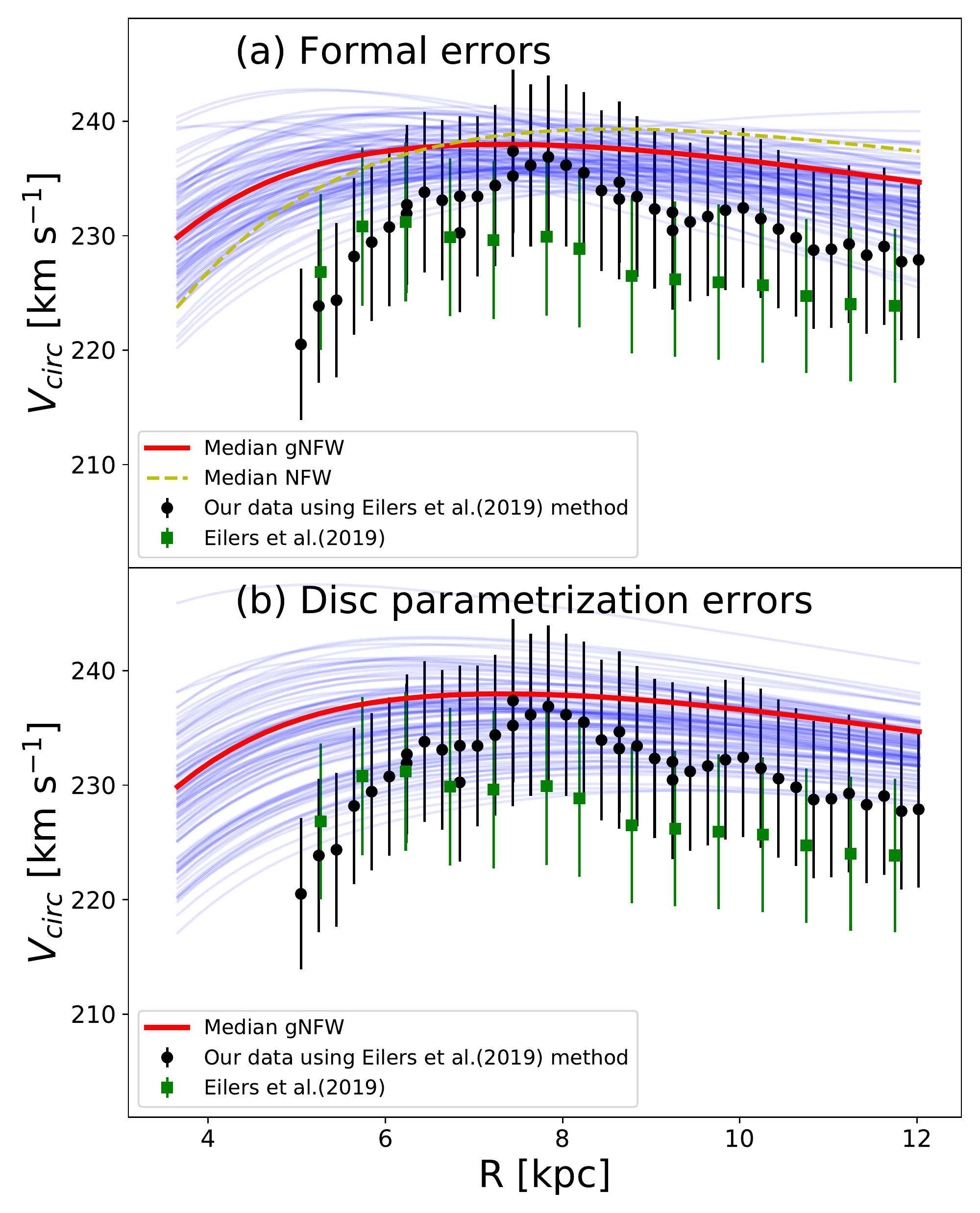}
    \caption{{\bf Circular Velocity for the model.} The red line is the circular velocity of our median gNFW model, the green squares are the recent measurements of \protect\cite{eilers18}, the black points are derived with a subset of our data, restricting them to the equatorial plane and the \protect\cite{eilers18} method and the yellow dashed line is for the mean normal NFW profile. The blue lines in panel (a) are the circular velocities of 100 random parameters of the posterior distribution that give the formal error from the fit and in panel (b) they are 100 different disc models that give the systematic error due to disc parametrization.}\label{im:vcirc}
\end{figure}

Having found a model that describes our \textit{Gaia} data we can also derive the circular velocity, using the \textsc{mge\_vcirc} procedure in the \textsc{jampy} package (see footnote \ref{note:jam}).

The circular velocity of the Milky Way disc region is plotted as a red line for the gNFW and as a dashed yellow line for the normal NFW profile in panel (a) of \autoref{im:vcirc}. The black points are derived with a subset ($|z|$ < 0.5~kpc or $\tan(z/R)<6^\circ$) from our data when using the same non-parametric method of \cite{eilers18}, also based on the axisymmetric Jeans equations, to calculate the circular velocity and the green squares are the measurements of \cite{eilers18}. The small offset between our values and the previous work by \cite{eilers18} is coming from our slightly higher value of the solar velocity because of the newest \cite{Abuter19} results for the distance to the Galactic Centre. The difference between our values using the method of \cite{eilers18} (black dots) and our result from the JAM$_{\rm sph}$ model (red line) is expected, since the data used for \cite{eilers18} method are only covering the equatorial plane of the Galaxy ($|z|$ < 0.5~kpc or $\tan(z/R)<6^\circ$) while for our model all the data reaching up to 2.5~kpc in $z$-direction were used. Moreover, also the different parametrization of the tracer distribution in \cite{eilers18} method and our makes a difference. We have investigated this further and can reproduce almost exactly their circular velocity curve if we use the same kinematics and the same tracer distribution of \citet{pouliasis17} with a spherical ($q_{\rm DM}$ = 1) NFW dark matter halo (Nitschai et al. in prep.). Reassuringly, the two circular velocities are consistent within the estimated quite small systematic uncertainties. The model with NFW halo would produce an inconsistent circular velocity, supporting the finding that this is excluded by the \textit{Gaia} data.

The blue transparent lines are 100 random realizations for the circular velocity from the posterior distribution that indicates the formal uncertainty for our mean circular velocity but include our approximate treatment for systematics by scaling the input kinematic errors. At the solar position we get a value of $v_{\rm circ}(R_{\odot})= 236.5\pm 1.8$~km~s$^{-1}$. This value also agrees well with the value of \cite{eilers18} ($229.0\pm 0.2$~km~s$^{-1}$), where they use a quite different stellar sample, stellar distance estimates and modelling assumptions, and with another recent work by \cite{mroz19} ($233.6\pm 2.8$~km~s$^{-1}$) in which they construct the rotation curve of the Milky Way using the proper motion and radial velocities from \textit{Gaia} for classical Cepheids.

The result for the effect of the uncertainties of the parametrization of the Milky Way stellar tracer distribution on the circular velocity can be seen in panel (b) of \autoref{im:vcirc}. Here the blue lines indicate the systematic error due to the Milky Way stellar tracer model. This gives us an alternative estimate of the value and systematic error $v_{\rm circ}(R_{\odot}) = 236.3 \pm 3.1$~km~$s^{-1}$, which is comparable to the previous estimate. 

 \section{Conclusion}\label{sec:conclusion}

The model presented is the first stringent test of the Newtonian equations of motion on galactic scales. It demonstrates that we already have a remarkably accurate knowledge of the mass distribution in our Milky Way and we can concisely describe the main average characteristics of the observed stellar kinematics with a minimal set of assumptions. The model also shows that the average kinematics of the Milky Way, outside of the bar region, can be well described by an axisymmetric and equilibrium model. The fact that we can well describe the Galactic kinematics, regardless of relatively minor deviations from equilibrium and axisymmetry \citep{Widrow12,Gaia2018,Antoja2018}, is consistent with observations of external galaxies, where such deviations are widespread, but models can still capture the average kinematic properties \citep{cappellari13} and recover circular velocity profiles to 10 per cent accuracy \citep{leung18}, even from data of much inferior quality. 

The \textit{Gaia} data are going to improve in accuracy with subsequent data releases. Moreover, when not relying entirely on parallactic distances, one can significantly increase the extent of the region where kinematics can be measured \citep{Wegg2019}. Dynamical models will soon be able to study our Galaxy's density distribution over larger distances. These models are starting to provide a description of the dynamics of the Milky Way at a level that is impossible to achieve in external galaxies. This is providing a key benchmark for our knowledge of galactic dynamics that will complement much less detailed studies of much larger samples of external galaxies.

The kinematics used in this paper and the MGE components can be found as Supplementary data in the online version.

\section*{Acknowledgements}

M.S. Nitschai warmly thanks the two scholarships, `DAAD-PROMOS-Stipendium' and `Baden-W\"urttemberg-STIPENDIUM', for the financial support during the stay abroad working on this project and the Astrophysics sub-department of the University of Oxford for the hospitality during that time. N.N. gratefully acknowledges support by the Deutsche Forschungsgemeinschaft (DFG, German Research Foundation) -- Project-ID 138713538 -- SFB 881 (`The Milky Way System', subproject B8).




\bibliographystyle{mnras}
\bibliography{references} 


\bsp	
\label{lastpage}
\end{document}